# Dynamic 3D shape measurement based on the phase-shifting moiré algorithm

Canlin Zhou [1*], Shuchun Si[1], XiaoLei Li[3], Zhenkun Lei[2], Yanjie Li[3], Chaorui Zhang[1]

*1 School of Physics, Shandong University, Jinan 250100, China*

*2 Department of Engineering Mechanics, Dalian University of Technology, Dalian 116024, China*

*3School of Mechanical Engineering, Hebei University of Technology, Tianjin, 300130, China*

*4 School of Civil Engineering and Architecture, University of Jinan, Jinan, 250022, China*

*Corresponding author: Tel: +8613256153609; E-mail address: canlinzhou@sdu.edu.cn

In order to increase the efficiency of phase retrieval,Wang et.al. proposed a high-speed moiré phase retrieval method.But it is used only to measure the tiny object. In view of the limitation of Wang's method,we proposed a dynamic three-dimensional(3D) measurement based on the phase-shifting moiré algorithm.First, four sinusoidal fringe patterns with a $\pi/2$ phase-shift are projected on the reference plane and acquired four deformed fringe patterns of the reference plane in advance. Then only single-shot deformed fringe pattern of the tested object is captured in measurement process.Four moiré fringe patterns can be obtained by numerical multiplication between the the AC component of the object pattern and the AC components of the reference patterns respectively. The four low-frequency components corresponding to the moiré fringe patterns are calculated by the complex encoding FT(Fourier transform) ,spectrum filtering and inverse FT.Thus the wrapped phase of the object can be determined in the tangent form from the four phase-shifting moiré fringe patterns using the four-step phase shifting algorithm.The continuous phase distribution can be obtained by the conventional unwrapping algorithm. Finally, experiments were conducted to prove the validity and feasibility of the proposed method. The results are analyzed and compared with those of Wang's method, demonstrating that our method not only can expand the measurement scope, but also can improve accuracy.

## 1. Introduction

Owing to the engineering requirements, the 3D shape measurement of dynamic objects is a widely concerning issue [1-3].Many scholars have carried out extensive studies. For example, Yoneyama et. al. [4] firstly began to measure the 3D shape of moving objects using linear sensor. Hu et. al. [5-6] measured the surface profile of moving objects by using a line-scan camera. Cao et. al. [7-9] used equivalent phase shifts with the relative motion of objects and a camera to measure the shape of the rotating and moving objects. Li et. al. [10] adopted a method similar to Cao's, but improved pixel matching accuracy using a layered modulation method. Lu et. al. [11] analyzed the reason for failure in the phase shift algorithm caused by the motion of the object, and proposed the motion compensation method . Cong et. al. [12] proposed a Fourier transform-assisted phase shifting method to reconstruct 3D shape of dynamic scenes, which estimates the motion using Fourier transforms,then formulates a motion-related model to fit the dynamic scenes more accurately. Wilm et. al. [13] constructed a phase shift model by using the phase matching method. Essentially, the above methods belong to the phase shift algorithm which require multiple fringe pattern images. Furthermore, all these methods need to calibrate the fringe image mismatch caused by the motion of the object. Lohry et. al. [14] proposed a dynamic 3D profilometry method using a fast binary dithered pattern projection. Lu et. al. [15] proposed a method based on modulation, which projects a cross grating and then determines the shape by modulation after the calibration of the relationship between the modulation and the object height. Zhong et. al. [16] proposed a method combining the fringe pattern projection and the binocular vision to realize the dynamic 3D shape measurement. Liu et. al. [17] used the crossed grating projection and differential method to measure the object's profile. Furthermore, Liu et. al. [18] made use of the color CCD and color composite fringes projection to complete the dynamic measurement. Nguyen et. al. [19] proposed a single frame composite fringe projection method. Zhu et. al. [20] used a high-speed DLP projector to project composite color fringes and a monochrome camera to capture the fringes in RGB channels. Salahieh et. al. [21] presented a single-shot multi-polarization fringe projection imaging technique . Zhang et.al. [22] proposed a spatial quasi-phase-shifting technique. Guan et. al. [23] proposed a composite coded grating method, and Mohammadi et. al. [24-25] proposed a moiré fringe phase shifting measurement method using a digital reference grating. These methods can be classified as single frame projection 3D profilometry. However, phase shifting profilometry commonly requires three or more images, hence its accuracy is easily affected by the phase-step error due to the movement of the object.

Therefore, a high-speed camera is required. Nonetheless, the hardware improvement cannot solve the problem at its root, thus this leads to a limitation on its application. Theory modeling and a compensation strategy on the phase-step error from the movement can reduce the effect of the object's motion, but it is not sufficient for all cases. Obtaining an equivalent phase-step by the relative motion between the object and the camera needs strict rectification. The method combining fringe pattern projection and binocular vision requires a complicated hardware system.

The single frame 3D reconstruction technique (such as the Fourier transform and composite coded grating method) can eliminate the phase-step error from the motion in dynamic measurement, but there are still issues in the stability and accuracy when using these methods. The single frame moiré retrieval method can automatically construct phase-shifting grids, but the elimination of high-frequency carrier fringes and the obtainment of moiré information are relatively complex and the accuracy is easily affected.

Recently, Wang et. al. [26] proposed a high-speed moiré-based phase retrieval method. According to our experiences, Wang's method has the following restriction: the phase of the tested object is limited to a range within (0, $\pi$), therefore, the object is not also too high.

The purpose of this study was to improve Wang's method further. Inspired by ref. [27], we have proposed a dynamic 3D shape measurement based on phase-shifting moiré algorithm. We tested the proposed method with experiments and compared it with Wang's method. The results demonstrated that the proposed method can expand the scope of Wang's method.

## 2. Theory

### 2.1 Wang's method

The details for Wang's method can be found in ref. [26]. What follows is a brief review of the method. It requires two fringe images with and without the measured object. The two images can be illustrated by Eq. (1) and Eq. (2), respectively.

$$I(x, y) = a + b\cos[k_x x + k_y y + \varphi(x, y) + \varphi_a(x, y)] \tag{1}$$

$$I_0(x, y) = a + b\cos[k_x x + k_y y + \varphi_a(x, y)] \tag{2}$$

In Eqs. (1) and (2), $x$ and $y$ are the spatial coordinates, $k_x$ and $k_y$ are the carrier frequency along the $x$-axes and $y$-axes, respectively, $\phi(x, y)$ represents the phase information of the measured object, and $\phi_a$ is the phase distribution induced by aberrations in the imaging system, $a$ and $b$ are the background intensity and the fringe contrast, respectively.

After removing the DC term, Eqs. (1) and (2) can be rewritten as

$$I' = b\cos(k_x x + k_y y + \phi(x, y) + \phi_a) \tag{3}$$

$$I_0' = b\cos(k_x x + k_y y + \phi_a) \tag{4}$$

Multiplying Eq. (3) by Eq. (4) produces Eq. (5).

$$\hat{I} = \frac{b^2}{2}\left\{\cos[\phi(x,y)] + \cos[2k_x x + 2k_y y + 2\phi_a + \phi(x,y)]\right\} \tag{5}$$

In Eq. (5), the low-frequency components contain the object's phase information. With a low-pass filter, the low-frequency components can be obtained as shown in Eq. (6).

$$I_{\cos} = \frac{b^2}{2}\cos[\phi(x,y)] \tag{6}$$

The normalization method[28-29] is used to process the Eq.(6). Thus, the following Eq. (7) can be obtained.

$$I_{\cos\_Normal} = \cos[\varphi(x,y)] \tag{7}$$

Finally, the phase distribution can be retrieved with arc cosine computing.

$$\varphi(x,y) = \arccos(I_{\cos\_Normal}) \tag{8}$$

As the range of the arccosine is limited to (0, $\pi$), Wang's method is only able to restore the phase of a thin object and will fail to measure a thick object.

*2.2 Our method*

Focusing on the existing issues, we proposed a dynamic 3D shape measurement method based on the phase-shifting moiré algorithm. The basic idea of our method is that we use four sinusoidal patterns with $\pi/2$ phase-shifting of the reference plane to obtain the phase of the object. The phase is expressed in a tangent form through numerical moiré calculations between the AC components of the fringe pattern on the object and the AC component of the four phase-step fringe patterns on the reference plane. Therefore, we determine the phase by the arctangent calculation.

The four sinusoidal fringe patterns with the phase shift of $\pi/2$ on the reference plane can be expressed by Eq. (9)-(12)

$$I_1(x,y) = a + b\cos[k_x x + k_y y + \phi_a(x,y)] \tag{9}$$

$$I_2(x,y) = a + b\cos[k_x x + k_y y + \phi_a(x,y) + \pi/2] \tag{10}$$

$$I_3(x, y) = a + b\cos[k_x x + k_y y + \phi_a(x, y) + \pi] \tag{11}$$

$$I_4(x, y) = a + b\cos[k_x x + k_y y + \phi_a(x, y) + 3\pi/2] \tag{12}$$

From the equations above, we can obtain the background intensity as below.

$$a = \frac{I_1 + I_2 + I_3 + I_4}{4} \tag{13}$$

The AC components of four phase-step fringe patterns on the reference plane can be described as Eq.(14)- (17) respectively.

$$I_1^R = I_1(x, y) - a = b\cos[k_x x + k_y y + \phi_a(x, y)] \tag{14}$$

$$I_2^R = I_2(x, y) - a = b\cos[k_x x + k_y y + \phi_a(x, y) + \pi/2] \tag{15}$$

$$I_3^R = I_3(x, y) - a = b\cos[k_x x + k_y y + \phi_a(x, y) + \pi] \tag{16}$$

$$I_4^R = I_4(x, y) - a = b\cos[k_x x + k_y y + \phi_a(x, y) + 3\pi/2] \tag{17}$$

Eq. (3) is multiplied by Eq.(14)-(17) respectively, thus, we can obtain the results below.

$$\hat{I}_1 = I' \times I_1^R = \frac{b^2}{2}\{\cos[2k_x x + 2k_y y + 2\phi_a + \phi(x, y)] + \cos[\phi(x, y)]\} \tag{18}$$

$$\hat{I}_2 = I' \times I_2^R = \frac{b^2}{2}\{\cos[2k_x x + 2k_y y + 2\phi_a + \phi(x, y) + \pi/2] + \cos[\phi(x, y) - \pi/2]\} \tag{19}$$

$$\hat{I}_3 = I' \times I_3^R = \frac{b^2}{2}\{\cos[2k_x x + 2k_y y + 2\phi_a + \phi(x, y) + \pi] + \cos[\phi(x, y) - \pi]\} \tag{20}$$

$$\hat{I}_4 = I' \times I_4^R = \frac{b^2}{2}\{\cos[2k_x x + 2k_y y + 2\phi_a + \phi(x, y) + 3\pi/2] + \cos[\phi(x, y) - 3\pi/2]\} \tag{21}$$

According to Eq. (18) - (21), it is easy to find that the low-frequency components are moire fringes with the phase shift of π/2, which contain the phase information of the object without the aberration and carrier phase and corresponding spectra always locate in the center of the spectrum plane. Meanwhile, the fundamental frequency is twice that of the original projected fringe, the probability caused by the spectrum overlapping between the zero frequency and fundamental frequency will become less than that with Fourier transform profilometry(FTP), therefore, we can easily obtain the low-frequency components with a low-pass filter.

In order to obtain the low-frequency components, $\hat{I}_1$、$\hat{I}_2$、$\hat{I}_3$ and $\hat{I}_4$ should be firstly Fourier transformed to estimate their spectra, respectively, Then we need to make inverse transform(IFT) to the filtered low-frequency components. The same processes including forward Fourier transform(FT), spectrum filtering and IFT have to be implemented many times. FT operation is inefficient. Here, we adopt the complex FT algorithm[30] to encode a synthetic pattern from $\hat{I}_1$ and $\hat{I}_2$ to simplify the process. The intensity of the synthetic fringe pattern can be described as:

$$\hat{I}_c = \hat{I}_1 + i \cdot \hat{I}_2 \qquad (22)$$

Where the symbol $i$ denotes imaginary unit.

There is the inverse relationship between the complex fringe pattern and its real and imaginary parts:

$$\hat{I}_1 = \frac{I_c + I_c^*}{2} \;;\; \hat{I}_2 = i \cdot \frac{-I_c + I_c^*}{2} \qquad (23)$$

Thus, the two-dimensional (2D) FT of $\hat{I}_1$ and $\hat{I}_2$ can be determined directly from the 2D FT of $\hat{I}_c$, can be written as follows:

$$FT\{\hat{I}_1\} = \frac{FT\{I_c\} + Flip\{FT(I_c)^*\}}{2} \qquad (24)$$

$$FT\{\hat{I}_2\} = j \cdot \frac{-FT\{I_c\} + Flip\{FT(I_c)^*\}}{2} \qquad (25)$$

Where $FT(x)$ denotes 2D Fourier transform of $x$, $Flip(x)$ denotes flipping the matrix $x$ along horizontal and vertical direction.

Therefore, it is enough to calculate only once the 2D FT of $\hat{I}_c$ to calculate the 2D FT of $\hat{I}_1$ and $\hat{I}_2$.

After the spectra of $\hat{I}_1$ and $\hat{I}_2$ are obtained, we can make low-pass filtering, inverse Fourier transform. Thus, the low-frequency components in Eq. (18) and (19) can be obtained as follow:

$$\hat{I}_1' = \frac{b^2}{2}\{\cos[\phi(x,y)]\} \qquad (26)$$

$$\hat{I}_2' = \frac{b^2}{2}\{\cos[\phi(x,y) - \pi/2]\} \qquad (27)$$

The similar procedures are applied to $\hat{I}_3$ and $\hat{I}_4$, thus we can get the low-frequency components in Eq. (20) and (21) as follow:

$$\hat{I}_3' = \frac{b^2}{2}\{\cos[\phi(x,y) - \pi]\} \qquad (28)$$

$$\hat{I}_4' = \frac{b^2}{2}\{\cos[\phi(x,y) - 3\pi/2]\} \qquad (29)$$

The wrapped phase of the object can be estimated from Eq. (26)-(29) using the four-step phase shifting algorithm by Eq.(30)

$$\tan[\phi(x,y)] = \frac{\hat{I}_4' - \hat{I}_2'}{\hat{I}_1' - \hat{I}_3'} \qquad (30)$$

The main stages of the proposed algorithm are summarized as follows:

(1) project and capture four phase-shifted sinusoidal fringe pattern on the reference plane.

(2) calculate the AC components of the four phase-shifted fringe pattern according to Eq.(14)- (17).

(3) project a sinusoidal fringe pattern on the tested object, acquire the corresponding deformed pattern.

(4) calculate the AC component of the deformed pattern on the object.

(5) calculate the moire fringe pattern according to Eq.(18) -(21).

(6) encode a synthetic pattern from $\hat{I}_1$ and $\hat{I}_2$ according to Eq.(22), apply FT to Eq.(22).

(7) obtain the spectra of $\hat{I}_1$ and $\hat{I}_2$ according to Eq.(24) -(25).

(8) obtain the low-frequency components as shown in Eq.(26) -(27) by making low-pass filtering and inverse Fourier transform.

(9) obtain the low-frequency components as shown in Eq. (28) -(29) after applying the similar procedures to $\hat{I}_3$ and $\hat{I}_4$.

(10) calculate the wrapped phase of the object from the low-frequency components as

shown in Eq. (26) -(29) according to Eq.(30).

(11) unwrap the wrapped phase by the phase unwrapping algorithm.

Eq.(30) shows that the normalization of the modulation term in Eq. (6) and(7) is omitted. Meanwhile, the phase of the object can be determined directly through the arctangent calculation. Thus, the conventional unwrapping algorithms[31-33] can be applied to unwrap the wrapped phase. Therefore, the proposed method eliminates the restriction that Wang's method can only measure thin objects.

## 3. Experiments and discussions

In this section, for evaluating the performance of proposed algorithm, we tested it in a series of experiments.

We developed a fringe projection measurement system, which consisted of a DLP projector (Optoma EX762) driven by a computer and a CCD camera (DH-SV401FM). The captured image is 768 pixels by 576 pixels. The software was programmed in Matlab with an i5-760 CPU at 2.80 GHz. We chose the object with a rough surface as the tested specimen, did not consider measuring objects with large reflectivity variations across their surface [34-35].

Firstly, an experiment is provided to demonstrate the feasibility of the proposed algorithm. The measured object is a simple mask with a large depth range. Initially,Wang's method was used to measure it. The captured fringe pattern of the reference plane is displayed in Fig. 1(a) , Fig. 1(b) presents the deformed fringe pattern of the mask. Figure 1(c) shows the result of moiré multiplication between Figs. 1(a) and 1(b) without the DC terms. Figure 1(d) displays the spectrum image of Fig 1(c). The moire fringe is shown in Fig. 1(f) after the low-pass filter as shown in Fig. 1(e). The phase of the object can be extracted by an arccosine calculation, which is shown in Fig. 1(g).

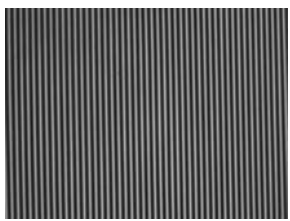 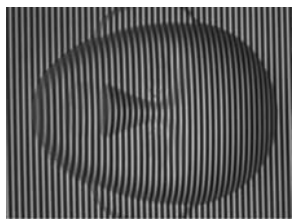 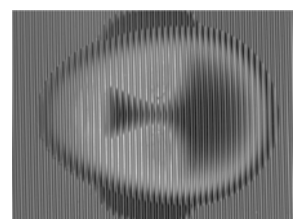

      (a)                       (b)                       (c)

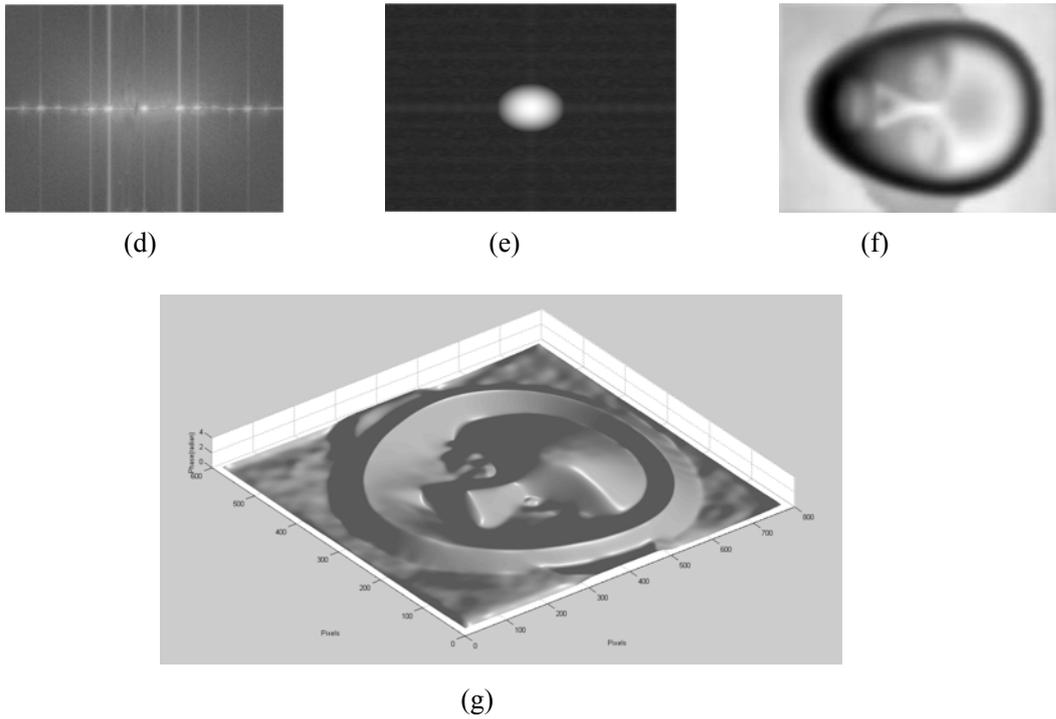

(d)                          (e)                          (f)

(g)

Fig. 1 (a) Fringe pattern of the reference plane;　　　(b) deformed　fringe pattern of the mask; (c) result of the moiré multiplication; (d) spectrum of (c); (e) low pass filter; (f) moiré fringe extracted from (c); (g) 3D shape map　restored　by Wang's method.

Then, we adopted proposed method to measure the mask. Four fringe patterns with a $\pi/2$ phase shift of the reference plane are captured in advance, which is shown in fig.2(a), their AC components are calculated according to Eq.(14)-(17). In measurement process, a deformed fringe pattern of the object is captured, which is shown in fig.2(b), its AC component is also calculated. The four moire fringe patterns are obtained according to Eq.(18) -(21). The low-frequency components corresponding to four moiré fringe patterns as described in Eq. (26) -(29) are estimated through the complex encoding FT, spectrum filtering and inverse FT, which are shown in Fig.2(c)-(f), Then the wrapped phase of the mask can determined by Eq.(30), After the phase unwrapping, the mask is restored effectively as shown in Fig.2(g).

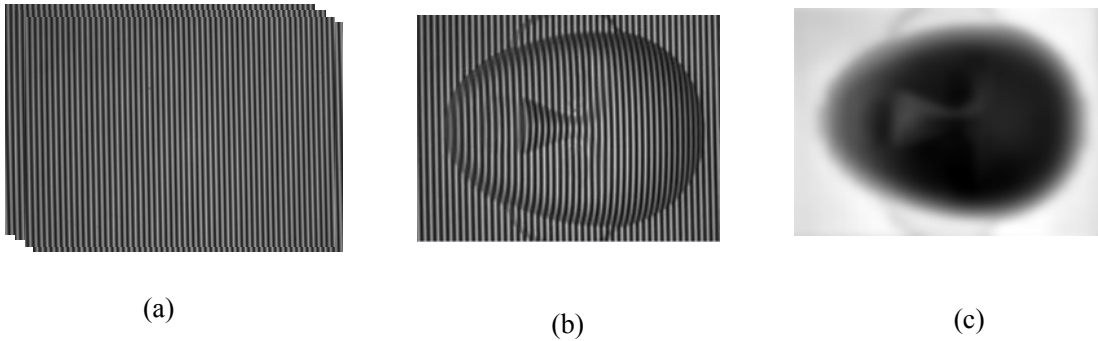

(a)                          (b)                          (c)

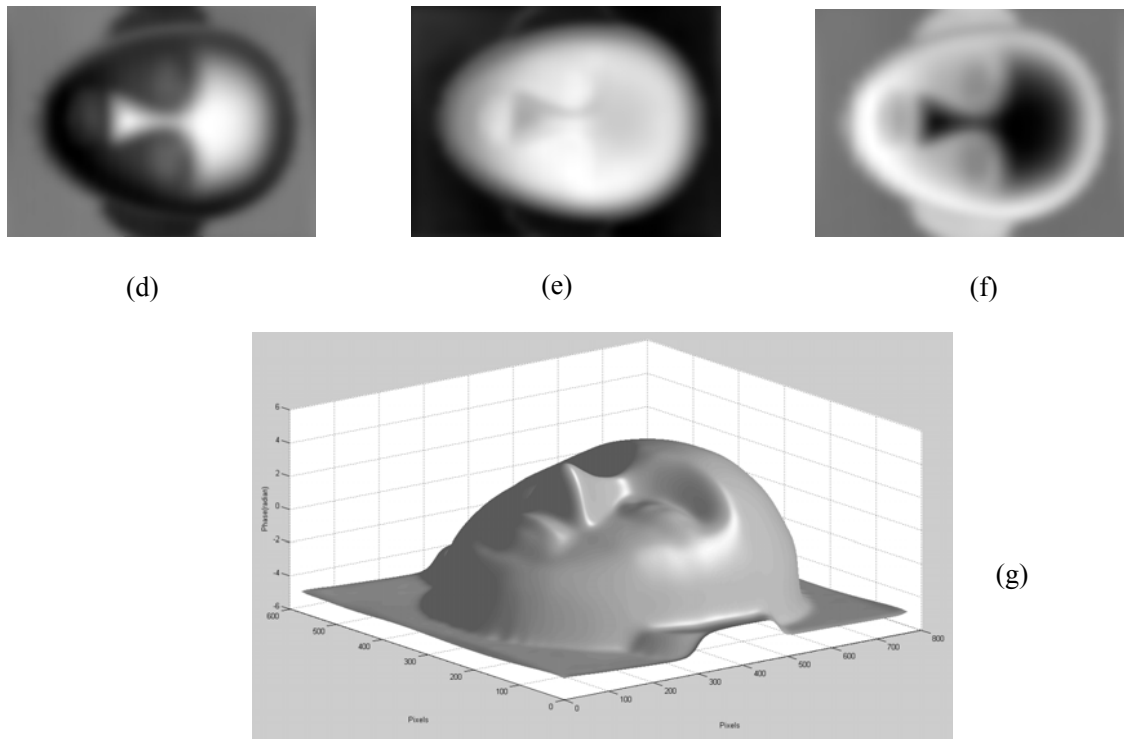

(d)　　　　　　　　　　(e)　　　　　　　　　　(f)

(g)

Fig. 2 (a) Four fringe patterns with a π/2 phase shift of the reference plane; (b) deformed　fringe pattern of the mask; (c) the low-frequency components　corresponding to 0-degree moiré fringe ; (d) the low-frequency components　corresponding to 90-degree moiré fringe　(e) the low-frequency components　corresponding to 180-degree moiré fringe (f) the low-frequency components　corresponding to 270-degree moiré fringe (g) 3D shape map　restored　by the proposed method.

　　By comparison of the results as shown in　Figs. 1(g) and 2(g), we can see that the proposed method retrieve the object well while Wang's algorithm can not do it .
　　Secondly, to show the repeatability of the proposed method, we do another experiment on two separated objects with complex shapes. The tested specimen are　two isolated objects, the face module and half of a paper cup, which have obvious change of height, are shown in Fig. 3.

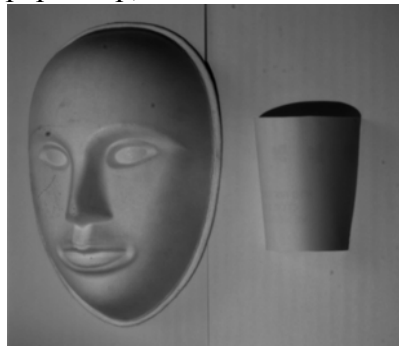

Fig. 3　the more complex face model and a half of paper cup cup.

　　The process of experiment is similar to the previous one. Fig.4 shows the captured images and results obtained by Wang's method. Fig. 4(a) represents the captured fringe pattern of the reference plane , Fig. 4(b) presents the deformed fringe pattern of the specimen. Figure 4(c) shows the moire fringe. 3D display of　the specimen's continuous phase is shown in Fig.4(d).

Fig. 5 gives the captured images and results processed by the proposed method. Four fringe patterns with a π/2 phase shift of the reference plane are captured in advance,which are shown in fig.5(a),their   AC components are calculated according to Eq.(14)-(17).In measurement process,a deformed fringe pattern of the object is captured,which is shown in fig.5(b), its AC component is also calculated. The four moire fringe patterns are obtained according to Eq.(18) -(21).The low-frequency components corresponding to  four  moiré fringe patterns as described in Eq. (26) -(29) are estimated through the complex encoding FT ,spectrum filtering and inverse FT,which are shown in Fig.5(c),Then the wrapped phase of the mask can determined by Eq.(30).After the phase unwrapping, 3D  display of  the specimen phase is shown in Fig.5(d).

Owing to the complexity of the specimen, there are some shadow areas in the captured fringes. They can be seen obviously from Fig. 4(b) and Fig. 5(b). From the   restored result as shown in Fig.4(d)   and   Fig.5(d), it is obvious that the proposed method works well to reconstruct the object, whereas Wang's method fails to restore the object.

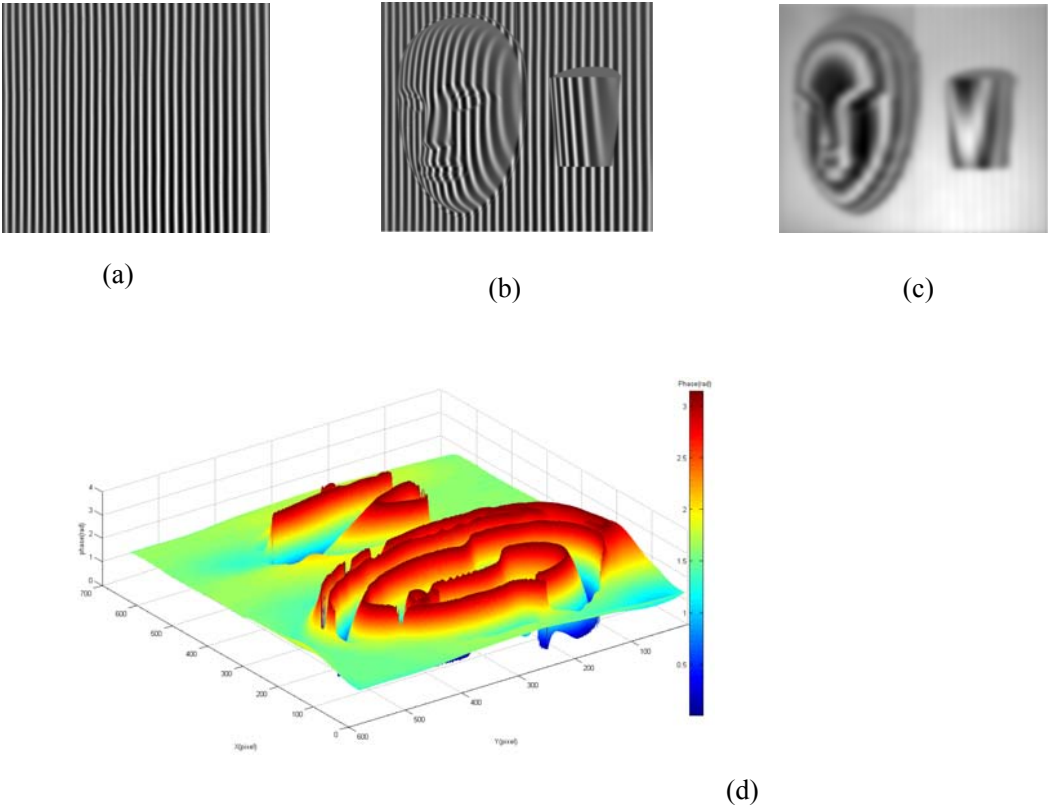

(a)  (b)  (c)

(d)

Fig. 4 (a) Fringe pattern of the reference plane; (b) deformed   fringe pattern of the specimen; (c) moiré fringe; (g) 3D shape map   of the specimen restored   by Wang's method.

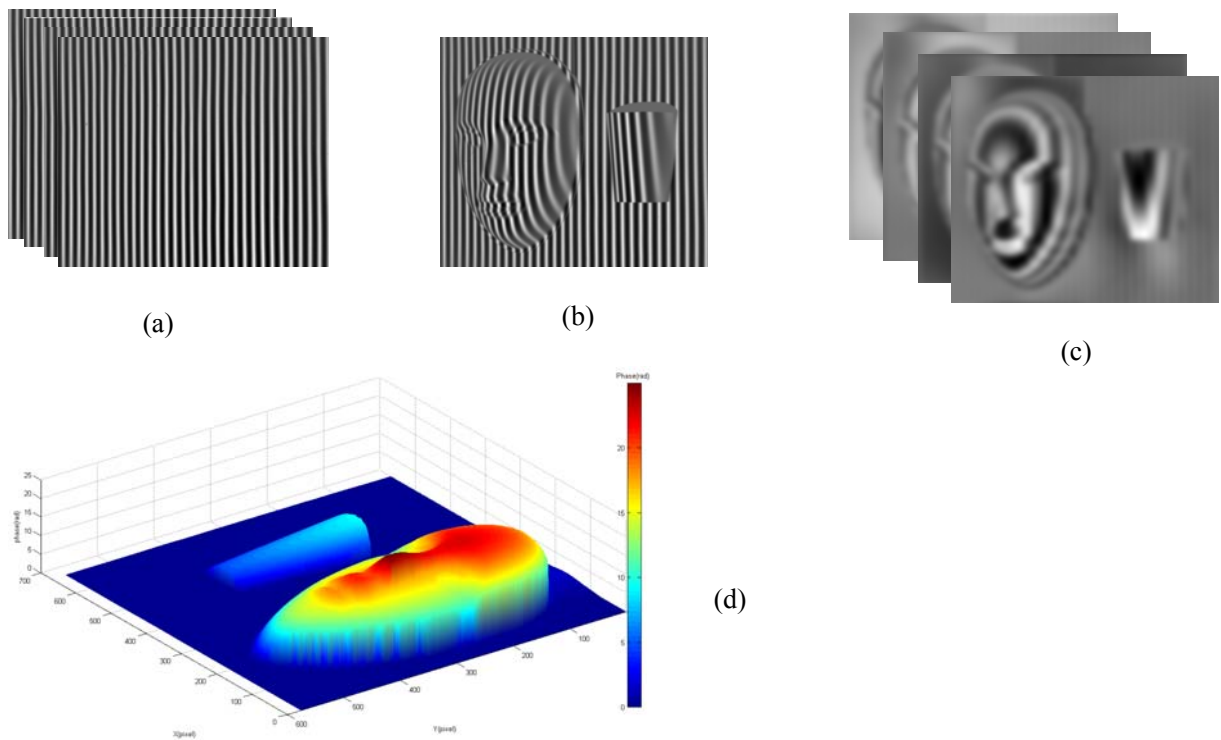

Fig. 5 (a) Four fringe patterns with a π/2 phase shift of the reference plane; (b) deformed fringe pattern of the specimen; (c) four low-frequency components corresponding to four phase-shifted moiré fringe ; (d) 3D shape map restored by the proposed method.

In order to evaluate the accuracy of the proposed method, we used the four-step phase shifting(FPS) method to obtain the phase map, which is used as the the baseline value for comparison. The root mean square error(RMS) of the results obtained by Wang's method and the proposed method are 1.9627 rad and 0.2528 rad respectively.The mean absolute error(MAE) of the results obtained by Wang's method and the proposed method are 2.5627 rad and 0.3162 rad respectively.It is clearly found that the proposed method has higher measurement accuracy compared to Wang's method.

## 4. Conclusion

In this paper, we proposed a dynamic 3D shape measurement method based on the phase-shifting moiré algorithm. It is an extension of Wang's method,which can expand its applicability to 3D measurement with a large depth range,improve the measurement accuracy. A series of experiments were carried out to verify the feasibility and validity of the proposed method. The experimental results demonstrate that the proposed method could accurately restore the 3D shape of the object with a large depth range and complex surface. Only one fringe pattern of the object was required to reconstruct the object after the four fringe patterns with a π/2 phase shift of the reference plane were captured in advance. Therefore, the proposed method is suitable for dynamic measurement or online measurement. The result of this research is valuable in dynamic 3D shape measurement.

Acknowledgment

This work was supported by the National Natural Science Foundation of China (NSFC) [Grant Nos: 11672162 and 11772081]. The support is gratefully acknowledged.